\documentclass[final,1p,times]{elsarticle}

\usepackage{amssymb}
\usepackage{lipsum}
\usepackage{amsthm}

\usepackage{amsfonts}
\usepackage{mathrsfs}
\usepackage{systeme, mathtools}

\usepackage{tikz-cd}
\usepackage{xcolor}
\usepackage{graphicx}% Include figure files
\usepackage{dcolumn}% Align table columns on decimal point
\usepackage{bm}% bold math
\usepackage{appendix}
\usepackage{url}
\usepackage{hyperref}
\hypersetup{hidelinks}
\numberwithin{equation}{section}
\newtheorem{theorem}{Theorem}[section]
\newtheorem{lemma}[theorem]{Lemma}
\newtheorem{proposition}[theorem]{Proposition}

\newtheorem{conjecture}[theorem]{Conjecture}
\theoremstyle{definition}
\newtheorem{remark}[theorem]{Remark}
\newtheorem{definition}[theorem]{Definition}

\newcommand{\R}{\mathbb{R}}

\newcommand{\M}{\mathcal{M}}
\newcommand{\q}{\hat{q}}
\newcommand{\p}{\hat{p}}
\newcommand{\tensor}{\otimes}

\newcommand{\iprod}{\mathbin{\lrcorner}}
\newcommand{\pder}[2]{\frac{\partial #1}{\partial #2}}

\journal{Nuclear Physics B}
\makeatletter
\def\ps@pprintTitle{%
 \let\@oddhead\@empty
 \let\@evenhead\@empty
 \let\@oddfoot\@empty
 \let\@evenfoot\@empty}
\makeatother

\begin{document}

\begin{frontmatter}



\title{Generalized Uncertainty Principle theories and their classical interpretation}


\author[Sap]{Matteo Bruno}
\ead{matteo.bruno@uniroma1.it}
\author[Sap,INFN]{Sebastiano Segreto}
\ead{sebastiano.segreto@uniroma1.it}
\author[ENEA,Sap]{Giovanni Montani}
\ead{giovanni.montani@enea.it}
\affiliation[Sap]{organization={Physics Department, Sapienza University of Rome},
            addressline={P.za Aldo Moro 5}, 
            city={Rome},
            postcode={00185}, 
            state={},
            country={Italy}}
\affiliation[ENEA]{organization={ENEA C.R.}, 
            addressline={Via E. Fermi 45}, 
            city={Frascati},
            postcode={00044}, 
            state={},
            country={Italy}}
\affiliation[INFN]{organization={INFN, Sezione               di Roma 1},
            addressline={P.le Aldo Moro 2}, 
            city={Rome},
            postcode={00185}, 
            state={},
            country={Italy}}

\begin{abstract}
 In this work, we show that it is possible to define a classical system associated with a Generalized Uncertainty Principle (GUP) theory via the implementation of a consistent symplectic structure. This provides a solid framework for the classical Hamiltonian formulation of such theories and the study of the dynamics of physical systems in the corresponding deformed phase space. \\
 By further characterizing the functions that govern non-commutativity in the configuration space using the algebra of angular momentum, we determine a general form for the rotation generator in these theories and crucially, we show that, under these conditions, unlike what has been previously found in the literature at the quantum level, this requirement does not lead to the superselection of GUP models at the classical level. \\
 Finally, we postulate that a properly defined GUP theory can be correctly interpreted classically if and only if the corresponding quantum commutators satisfy the Jacobi identities, identifying those quantization prescriptions for which this holds true.

\end{abstract}




\end{frontmatter}

%\tableofcontents



\section{Introduction}
\label{introduction}

 Generalized uncertainty principle (GUP) theories are quantum non-relativistic models based on a modification of the usual Heisenberg uncertainty principle (HUP), capable of providing an effective framework to account for an altered structure of space-time.

 Several arguments, stemming from String Theory and Gedanken experiments \cite{Amati:1988tn, Konishi:1989wk, Maggiore:1993rv}, suggest that a modification of the HUP is indeed necessary at some level. It is straightforward to realize that the most rigorous way to implement such a modification is to properly induce a different structure of the Heisenberg algebra, i.e. alter the canonical commutation relations between quantum conjugate operators. According to the type of modification introduced in the algebra, different classes of GUP theories can be identified. In the class of interest in this paper, whose structure will be discussed below, the main possible physical consequences due to the constructed deformation are twofold: first, the appearance of a minimum structure in the configuration space of the theory, e.g. a minimum length in position space, in the form of a non-zero minimal uncertainty in the configuration operators; second, a resulting non-commutativity between the configuration operators, which can be interpreted as the emergence of a non-commutative ``geometry" in configuration space \cite{Kempf:1994su, Kempf:1996nk, Segreto:2022clx, Segreto:2024vtu}.

 Being effective models, GUP theories are characterized by a deformation parameter, usually denoted as $\beta$, which naturally sets the scale of the deformation of the algebra or, in other words, the energy regime in which the physical effects described above become potentially relevant.

 Both of these features address the need to reconsider the space-time structure at some level, a common and widespread idea in any theory of quantum gravity, and in this spirit, it is clear that the scale of corrections provided by GUP theories is a Planckian one.

 In this respect, even if applications of GUP theories can be found in countless physical scenarios \cite{Bosso:2023aht}, perhaps the most relevant implementation of these effective models concerns early-stage cosmology \cite{Battisti:2007jd, Battisti:2007zg, Battisti:2008rv} and black hole physics \cite{Bosso:2023fnb, Ong:2018zqn}. In these contexts, they can introduce a drastically different dynamics at the Planck scale, providing useful physical insights on the role played by quantum gravity.

 From all these considerations, it is well understood that the natural formulation of GUP theories is in the quantum setting. Nevertheless, in the literature, it is possible to find numerous works in which the GUP framework is implemented at the classical level \cite{Battisti:2008qi, Barca:2021epy}. To achieve this, the deformed structure of the commutators is inherited by the Poisson brackets, which dictate the Hamiltonian dynamics. From a physical point of view, the essence of these studies is clearly a semi-classical formulation of the dynamics of the systems of interest: the classical trajectories acquire some corrective terms, which should be relevant at some level, arising in principle from the underlying quantum theory.

 Despite the soundness of the physical approach, from a formal perspective, given a set of commutators, it is not an obvious matter to determine if these are compatible with the classical Hamiltonian formulation of the dynamics.

 Hamiltonian classical mechanics can be completely recast in the powerful language of symplectic geometry. In this formulation, the phase space is represented by an even-dimensional smooth manifold $\mathcal{M}$ equipped with a \emph{non-degenerate} and \emph{closed} differential \emph{2}-form, called \emph{symplectic form}, which will induce a Poisson structure on $\M$ \cite{Libermann_1987, Lee_2003}. Once these objects are constructed coherently, the Hamiltonian dynamics we are familiar with can be correctly obtained and used as a framework to study the system of interest in phase space.
 The definition of such a \emph{2}-form with these precise requirements is the key ingredient to achieve a correct Hamiltonian formulation of the classical dynamics. More specifically, once we have equipped our phase space with a symplectic form and hence induced a Poisson structure, we can define a Hamiltonian vector field associated with a preferred smooth function $H$ called Hamiltonian function. The non-degeneracy of the \emph{2}-form ensures that the evolution of the Hamiltonian vector field is uniquely determined by the corresponding Hamiltonian $H$; the closure, on the other hand, allows us to preserve the symplectic form - hence the Hamiltonian dynamics - along the Hamiltonian vector field itself.
 Clearly, not all Hamiltonian systems are supposed to respect these conditions. For example, gauge systems should be described by degenerate \emph{2}-forms and non-holonomic systems by a \emph{2}-form not-closed. 
 These cases will not be relevant for our purpose.

 In light of this, in this paper, we investigate if it is possible to correctly implement a GUP theory at the classical level, using the symplectic geometry formulation. 
 Given an even-dimensional smooth manifold $\mathcal{M}$, we interpret the variables $(q,p)$ as coordinates on $\mathcal{M}$ and define a \emph{2}-form which induces the desired GUP Poisson brackets, requiring this \emph{2}-form to be symplectic, i.e., non-degenerate and closed. This allows us to completely characterize the symplectic structure of the theory in order to be consistent, finding precise and completely general relations between the phase-space functions which control the deformation of the Poisson structure. 
 Consistently, the results we obtain are entirely equivalent to imposing the validity of Jacobi identities for the deformed Poisson brackets, exactly as it should be.

 On the basis of these general symplectic relations, we further constrain the Poisson brackets between the generalized coordinates. Specifically we require that they form a set of phase-space functions, whose algebra corresponds to that of an angular momentum. This request, commonly imposed in many GUP theories, helps to maintain the rotational invariance of the algebra. 
 This allows us to further specify the structure of the Poisson brackets and determine the generator of rotations in these theories. Our results are essentially compatible with those of \cite{Maggiore:1993kv, Fadel:2021hnx}, where a similar analysis is conducted at the quantum level using the Jacobi identities. However, while \cite{Maggiore:1993kv, Fadel:2021hnx} suggests that the angular momentum operator constrains the theory additionally due to the presence of the spin, our analysis shows that this possibility is ruled out at the classical level. Specifically, we demonstrate that an angular momentum, which proper generates the rotations, exists as a phase-space function but, since no analogue of the spin emerges classically, does not impose any other constraint.\\
 As a primary consequence, this implies, as we will discuss in detail, that at the classical level no specific GUP model is favored, in contrast to what is argued in the quantum case in \cite{Maggiore:1993kv, Fadel:2021hnx}.

 The complete general determination of the symplectic structure of the classical theory ultimately allows us to formulate a Conjecture that concludes this work: a GUP theory is consistently implemented at the classical level if and only if the corresponding quantum commutators satisfy the Jacobi identities.\\
 Indeed, if the Jacobi identities are satisfied at the quantum level, it is straightforward to show that they will also be satisfied at the classical level. Due to the equivalence between the validity of the Jacobi identities and the non-degeneracy and closure of the symplectic form, the classical theory constructed in this way will be well-defined, with a Poisson structure related to the symplectic form. \\
 Conversely, if the classical relations that ensure the theory is symplectic are satisfied, meaning that the classical Jacobi identities hold, it is not obvious that the quantum Jacobi identities are satisfied due to the ambiguities in the quantization procedures. Nevertheless, we show that some of the most natural ordering prescriptions satisfy the Jacobi identities, thereby providing a potential guideline for the prescription of the theory at the quantum level.

 The paper is organized as follows: Section \ref{Sec:ClasInter} reviews fundamental elements and concepts of symplectic geometry relevant to the formulation of Hamiltonian classical mechanics. Moreover, we introduce the concept of \textit{classical interpretation} of a GUP theory and the main Conjecture associated with this definition.
 
 In Section \ref{Sec:build_f}, we derive the symplectic structure of a general GUP theory within the class of interest, based on its Poisson brackets. We particularly focus on the relationships dictated by the closure condition of the \emph{2}-form. We perform computations ranging from \emph{1} to \emph{d}-dimensional cases, providing concrete examples of GUP theories that align perfectly with the derived relations, including, e.g. the GUP model found in \cite{Kempf:1994su}, along with explicit calculations.
 
 In Section \ref{Sec:Maggiore} we consider the case where an angular momentum function defines the Poisson brackets between generalized coordinates and we suitably compare our findings with those presented in \cite{Maggiore:1993kv, Fadel:2021hnx}.\\ 
 In Section \ref{Sec:Proof} we present a wide class of cases for which the Conjecture presented in Section \ref{Sec:ClasInter} holds.  
 The Conjecture provides a relation between the classical and quantum framework of these theories.\\
 Finally, in  Section \ref{Sec:Concl} we give our conclusions, emphasizing the significance of clarifying the consistency of GUP theories at the classical level for any semiclassical applications.
 
 Additionally, an appendix, \ref{appendix}, in which we review the relation between the Poisson brackets and the Jacobi identities is included.

\section{Classical interpretation and Jacobi identity}
\label{Sec:ClasInter}
 The class of GUP theories of which we aim to discuss the classical interpretation is the one described by the following commutation relations:
  \begin{align}
  \label{GUP}
    \nonumber
    &[\hat p_i,\hat p_j]=0,\\
    &[\hat q_i,\hat q_j]=i\hbar L_{ij}(\hat q,\hat p),\\ \nonumber    
    &[\hat q_i,\hat p_j]=i\hbar\delta_{ij} f(\hat q,\hat p).
  \end{align}
  
 By \emph{classical interpretation} of a \emph{d}-dimensional GUP theory we can give the following definition:
 \begin{definition}
 \label{class}
    The classical interpretation of a GUP theory in the class given above consists of interpreting the set $(q_1,\dots,q_d,p_1,\dots,p_d)$ as coordinates on a \emph{2d}-dimensional smooth manifold $\M$ equipped with a symplectic form $\omega$ such that it induces the following fundamental Poisson brackets
    \begin{align}
    \label{Poisson}
    \nonumber
    &\{ p_i, p_j\}=0,\\
    &\{ q_i, q_j\}= L_{ij}(q,p),\\ \nonumber  
    &\{ q_i,p_j\}=f(q,p)\delta_{ij}.
    \end{align}
    
    We refer to the symplectic manifold $(\M,\omega)$ as phase space.
 \end{definition}
 
 \begin{remark}
   A symplectic form $\omega$ induces a Lie algebra structure on $\mathcal{C}^{\infty}(\M)$ via the Poisson brackets. Let $f,g\in\mathcal{C}^{\infty}(\M)$, and let $X_f$ be Hamiltonian vector field of $f$ defined by $df=X_f \iprod \omega$, the Poisson brackets of two functions are $\{f,g\}=-\omega(X_f,X_g)$. The Poisson brackets satisfy the Jacobi identity and define a non-degenerate Poisson structure on $\M$ \cite{Lee_2003}.
   
 \end{remark}
 From Definition \ref{class} we can compute the matrix associated with the symplectic form that induces the Poisson brackets \eqref{Poisson} in the coordinate system $x^a=(q_1,\dots,q_d,p_1,\dots,p_d)$. Indeed, the inverse of the symplectic form matrix is easily computable:
 \begin{equation}
 \label{PoissonMat}
   \omega^{ab}=\{x^a,x^b\}=
   \begin{pmatrix}
    L & f\mathrm{id}_d\\
    -f\mathrm{id}_d & 0
   \end{pmatrix},
 \end{equation}
 The inverse of this matrix is immediate and so we obtain the matrix of the symplectic form
 \begin{equation}
 \label{Sp-form}
  \omega_{ab}=
  \begin{pmatrix}
    0 & -\frac{1}{f}\mathrm{id}_d\\
    \frac{1}{f}\mathrm{id}_d & \frac{1}{f^2}L
  \end{pmatrix},
 \end{equation}
 where clearly $L:=\{L_{ij}\}$ is a skew-symmetric submatrix and $\mathrm{id}_d$ is the identity matrix in $d$ dimensions.

 From \eqref{Sp-form} it is straightforward to realize that, if we want to deal with a properly defined smooth $\emph{2}$-form, we have to require $f(x)\neq0,\,\forall x\in \M$. Hence, $f$ is a function with a definite sign all over $\M$.
 Specifically, in order to obtain ordinary classical mechanics in a proper limit, these functions have to be strictly positive functions.
 Usually, this limit is managed by introducing some suitable parameter that controls the deformation of the Poisson structure. As this parameter approaches zero, ordinary Poisson brackets are fully recovered.\\\\
 However, the \emph{2}-form in \eqref{Sp-form} is not a symplectic form \textit{a priori}. This classical structure is a relevant feature for the study of quantum physical theory. Indeed, despite in the literature it is debated if the semiclassical dynamics of a GUP is implemented by its classical interpretation, we can anyway relate the algebraic properties of the quantum commutation relations to the symplectic structure. When $L_{ij}$ and $f$ can be interpreted as smooth functions of the coordinates on an open subset $\mathcal U\subset \R^{2d}$, and $f$ is strictly positive on $\mathcal U$, we can formulate the following Conjecture:
 \begin{conjecture}
 \label{Theo}
    A \emph{d}-dimensional GUP theory as in \ref{GUP} admits a classical interpretation defined in Definition \ref{class} if and only if the commutators satisfy the Jacobi identity.
\end{conjecture}
In Section \ref{Sec:Proof} we provide a series of cases in which the Conjecture is verified. 
As we will see, we can postulate that the Conjecture provides us with a correct prescription for the quantization of some relevant phase-space functions, selecting suitable operator orderings among all the possible ones.\\
The following Section shows how to construct the symplectic form $\omega$ just imposing the \emph{non-degeneracy} and the \emph{closure} of $\omega$, funding some constraints for the expressions of $L$ and $f$. The considerations expressed in Section \ref{Sec:build_f} will constitute the main part of the construction of the Conjecture \ref{Theo}.\\\\
 The non-degeneracy condition is quite trivial since $\det (\omega_{ab})=f^{-2d}$. Therefore, to fulfill the non-degeneracy request it suffices to deal with smooth strictly positive function $f$ on $\M$, as already required.\\
 The closed condition, on the other hand, implies that we have to verify that $d\omega=0$.
 As we are going to see, this condition will give us a complete set of equations for $L$ and $f$ and it will establish a set of relations between these two objects.

\section{Building the function f(p)}
 \label{Sec:build_f}
 In this section, we are going to explicitly compute the equations coming from $d \omega=0$, for the \emph{1}, \emph{2}, and finally \emph{d}-dimensional case.
 Furthermore, we will provide some explicit examples and we will discuss, within this symplectic framework, some well-known GUP models scattered in the literature.
 
 \subsection{\emph{1}-dimensional GUP theories}
 \label{1-d}
 The \emph{1}-dimensional case is highly trivial. 
 Being, in this case, the phase space a \emph{2}-dimensional manifold, every \emph{2}-form is closed.
 This means that the closed condition is automatically satisfied.
 From this, immediately follows that every \emph{1}-dimensional GUP theory, with a suitable $f$ respecting the minimal conditions imposed above, has a symplectic structure and hence a classical interpretation.

 \subsection{\emph{2}-dimensional GUP theories}
 \label{2-d}
  In the \emph{2}-dimensional case, the matrix $L_{ij}$ has a unique degree of freedom $L_{12}(q,p)=:l(q,p)$. Let us call $f^{-1}=h$. The symplectic form matrix now reads:
 \begin{equation}
 \label{2d-Sp_form}
    \omega_{ab}=
    \begin{pmatrix}
    0 & 0 & -h & 0\\
    0 & 0 & 0 & -h\\
    h & 0 & 0 & h^2l\\
    0 & h & -h^2l & 0
   \end{pmatrix}.
 \end{equation}
 The closure condition, namely $\partial_c\omega_{ab}dx^c\wedge dx^a\wedge dx^b=0$, gives us $\binom{4}{3}=4$ scalar equations, i.e. the components of the \emph{3}-form $\partial_c\omega_{ab}dx^c\wedge dx^a\wedge dx^b=2(\partial_c\omega_{ab}+\partial_a\omega_{bc}+\partial_b\omega_{ca})dx^c\tensor dx^a \tensor dx^b=0$.
 Explicitly, we obtain the following set of equations:
 \begin{equation}
 \label{2_dim_eqs} 
    \begin{cases}
        (123) & \pder{h}{q_2}=0,\\
        (124) & -\pder{h}{q_1}=0,\\
        (134) & \pder{h^2l}{q_1}-\pder{h}{p_2}=0,\\
        (234) & \pder{h^2l}{q_2}+\pder{h}{p_1}=0.
    \end{cases}
 \end{equation}
 The first two say that $h$, hence $f$, does not depend on $q_1,q_2$, but only on the momenta variables $p_1,p_2$.
 Therefore, considering $h=f^{-1}(p_1,p_2)$, the last two equations of \eqref{2_dim_eqs} can be rewritten as:
 \begin{equation}
 \label{2d_grad_eq}
    \pder{l}{q_1}+\pder{f}{p_2}=0,\ \ \ \pder{l}{q_2}-\pder{f}{p_1}=0.
 \end{equation}
 These relations can be read as a gradient equation for $f$ in an open set $U \subset \R^2 $ with respect to $p_1,p_2$ variables.
 Together with the independence of $f$ from $q_1,q_2$, from \eqref{2d_grad_eq} we can deduce a general form for the $l$ function:
 \begin{equation}
  \label{2d_l_form}
     l(q,p)=s(p)+a(p)q_1+b(p)q_2,
 \end{equation}
 where $s,a,b$ are smooth functions of $p_1,p_2$.
 If $U$ is simply connected, the system \eqref{2d_grad_eq} admits solution if and only if
 \begin{equation}
 \label{2d_nec_suff_cond}
  \pder{a(p)}{p_1}=-\pder{b(p)}{p_2}.
 \end{equation}
 
 In this case, the solution for $f$ is given by the integral along a curve $\gamma\subset U$ with final point $(p_1,p_2)$ modulo an integration constant:
 \begin{equation}
  \label{2d_f_gen_sol}
     f(p_1,p_2)=\int_{\gamma} (\pder{l}{q_2},-\pder{l}{q_1})\cdot \dot \gamma(t) dt.
 \end{equation}
 Note that, in this setting, the system \eqref{2d_grad_eq} can be equivalently read as a gradient equation for $l$.

 \subsubsection{2D classical mechanics and 2D non-commutative classical mechanics}
 \label{l=k}
 
 Fixing $l=\kappa$, where $\kappa$ is an arbitrary constant, we find a trivial solution for $f$. The two equations \eqref{2d_grad_eq} impose indeed $f=const$.
 With a proper choice of the constants involved, here we can recover both ordinary $2$D classical mechanics ($\kappa$=0 and $f$=1) and one of the most studied formulations of $2$D non-commutative classical mechanics, which does not involve a modification of the Poisson brackets between conjugate variables (see e.g. \cite{Gamboa:2000yq}).

 \subsubsection{Interesting example: Kempf-Mangano-Mann 2D-GUP}
 \label{l_angular}
 Consider $l(q,p)$ proportional to the usual angular momentum $l=\kappa(q_1p_2-q_2p_1)$. This function is clearly compatible with the general form \eqref{2d_l_form} and satisfies the integrability condition \eqref{2d_nec_suff_cond}.\\
 In this case the gradient equation \eqref{2d_grad_eq} for $f$ assumes a simple form:
 \begin{equation*}
    \kappa p_2+\pder{f}{p_2}=0,\ \ \ \kappa p_1+\pder{f}{p_1}=0.
 \end{equation*}
 The solution is $f(p_1,p_2)=-\frac{\kappa}{2}(p_1^2+p_2^2)+const$, which essentially represents, modulo the factor $\kappa$, a kinetic energy term.\\
 The integration constant can be fixed by introducing a control parameter and requiring that, in a certain limit, one recovers the ordinary Poisson brackets. In the case taken into the analysis, this can be done by evaluating the limit for $\kappa \to 0$, which fixes the integration constant to be $1$.\\ 
 Clearly we are not considering a possible dependence of the integration constant on the control parameter $\kappa$.
 The idea is indeed to introduce the deformation parameter in a natural way only once and let it properly propagate along the equations. \\
 If we now fix $\kappa=-2 \beta$, where $\beta$ is a positive parameter, what we obtain is the well-known GUP model first proposed by Kempf, Mangano, and Mann in \cite{Kempf:1994su}, at the quantum level.\\
 Note that in order to deal with a strictly positive $f$, for the reason discussed above, either $\kappa<0$ or, for $\kappa>0$, our phase space is a ball in the space of momenta $\frac{\kappa}{2}(p_1^2+p_2^2)<1$.
 
 It is worth to notice that the symplectic form naturally induces a volume form in the phase space, which in the \emph{2}-dimensional case reads as $\mathrm{Vol}_{\omega}:=\omega \wedge \omega=2! f^{-2} \ dq^1\wedge dq^2 \wedge dp^1 \wedge dp^2$.
 
 This volume form is clearly well-defined if and only if the $\omega$ is non-degenerate, that is if $f \neq 0$. The resulting condition is very close to the one obtained in the quantum framework from the request of symmetry of the fundamental operators, in particular the position operator.
 In this context in order to recover the symmetry of the operators - necessary but not sufficient condition for the self-adjointness - we need to deform the Lebesgue measure in a very similar way, introducing a factor $1/f$.
 As a consequence, both measures result to be well-defined if and only if $f \neq 0$, which in turn is a condition related, in the classical context, to the non-degeneracy of $\omega$.
 
 \subsubsection{Explicit computation: Polynomial functions}
 \label{l_poly}
 In this Section, we aim to derive an explicit and general form for $f$ from a wide class of functions for $l$. We consider a polynomial function $l$:
 \begin{equation}
    l(q,p)=\sum \alpha_{rsmn}q_1^rq_2^sp_1^mp_2^n.  
 \end{equation}
 We already know that $l(q,p)$ can be a polynomial with maximal degree one with respect to $q$ variables. Given that, we can write:
 \begin{equation}
  l(q,p)=\sum\alpha_{mn}p_1^mp_2^n+q_1\sum\beta_{mn}p_1^mp_2^n+q_2\sum\gamma_{mn}p_1^mp_2^n,
 \end{equation}
 The necessary and sufficient condition \eqref{2d_nec_suff_cond} now reads as: 
 \begin{equation}
   \label{int_cond}
    m\beta_{m,n-1}+n\gamma_{m-1,n}=0,\ \forall m,n\in\mathbb{N}.   
 \end{equation}

 From the constraint (\ref{int_cond}), the polynomial $l$ has the following form: 
 \begin{align}
 \label{pol_l}
 \nonumber
    l(q,p)=&\sum_{m,n\geq 0}\alpha_{mn}p_1^mp_2^n+\sum_{m\geq 1,\,n\geq 0}\beta_{mn}p_1^mp_2^nq_1+\sum_{n\geq 0}\beta_{n}p_2^nq_1+\\
    &-\sum_{m\geq 0,\,n\geq 1}\tfrac{m+1}{n}\beta_{m+1,n-1}p_1^mp_2^nq_2-\sum_{m\geq 0}\gamma_{m}p_1^mq_2,
 \end{align}
 with $\beta_n\equiv \beta_{0n}$ and $\gamma_m\equiv -\gamma_{m0}$. \\
 The $f$ function, by reason of \eqref{2d_f_gen_sol}, can be computed on a simple curve
 \[\gamma(t)=\begin{cases}
     (t,0) & \quad \text{if } 0\leq t \leq p_1,\\
     (p_1,t-p_1) & \quad \text{if } p_1\leq t \leq p_2+p_1,
 \end{cases}\]
 resulting in
 \begin{equation*}
    f(p_1,p_2)=\int_{0}^{p_1}\pder{l}{q_2}(p_1',0)dp_1'-\int_{0}^{p_2}\pder{l}{q_1}(p_1,p_2')dp_2'+c.
\end{equation*}
From this solution, we can give an explicit form of $f(p)$ in the polynomial case.
 \begin{equation}
    \label{pol_sol}
    f(p_1,p_2)=-\sum_{m\geq 0}\tfrac{\gamma_{m}}{m+1}p_1^{m+1}-\sum_{n\geq 0}\tfrac{\beta_{n}}{n+1}p_2^{n+1}-\sum_{m,\,n\geq 1}\tfrac{1}{n}\beta_{m,n-1}p_1^{m}p_2^n+c.
 \end{equation}
 We can check \textit{a posteriori} that $f$ is smooth and, if $c>0$, there exists $\varepsilon>0$ such that $f$ is strictly positive in $\mathbb{B}_{\varepsilon}^2\subset\R^2$. Thus, our phase space is $\mathcal{M}=\R^2\times\mathbb{B}_{\varepsilon}^2$ with a symplectic form given in coordinates by \eqref{2d-Sp_form}, where the functions $l$ and $f$ are defined in \eqref{pol_l} and \eqref{pol_sol}, respectively.\\
 Note that the ``angular momentum" case is included in this description. It can be recovered by setting $\beta_1=\kappa=\gamma_1$ and all others coefficients null.

 \subsection{\emph{d}-dimensional GUP theories}
 \label{d-d}
 We can extend this analysis to GUP theories in every dimension \emph{d}. Analogously to the \emph{2}-dimensional case, from the closed condition of the \emph{2}-form in \eqref{Sp-form}, we find a set of differential equations for $L$ and $f$.\\
 Notice that we can rewrite the \emph{2}-form as $\omega=h\Tilde{\omega}$, where $h=f^{-1}(q,p)$ and
\[\Tilde{\omega}_{ab}=\begin{pmatrix}
    0 & -\mathrm{id}_d\\
    \mathrm{id}_d & hL
    \end{pmatrix}.\]
In coordinates $x^a=(q_1,\dots,q_d,p_1,\dots,p_d)$, the closure condition $d\omega=0$ reads
\begin{equation}
    \Tilde{\omega}_{ab}\partial_cf+\Tilde{\omega}_{bc}\partial_af+\Tilde{\omega}_{ca}\partial_bf -f \partial_c\Tilde{\omega}_{ab}-f \partial_a\Tilde{\omega}_{bc}-f \partial_b\Tilde{\omega}_{ca}=0.
\end{equation}
The equations can be collected in four sets, depending on the values of the indices. Clearly, the equations are symmetric under permutations of the indices, obtaining at most $\binom{2d}{3}$ independent equations.\\\\
For $a,b,c\leq \emph{d}$, we have a set of trivial identities. So, that choice of indices is not relevant.\\\\
For $c>d$ and $\,a,b\leq d$, defining $k=c-d$ and $i,j \in\{1,d\}$, we get:
\begin{equation}
\label{d-q_indip}
    -\pder{f}{q_i}\delta_{jk}+\pder{f}{q_j}\delta_{ik}=0.
\end{equation}
That is the independence of $f$ on $q$s. We can show it by noticing that the equations are antisymmetric in $i,j$ and the only non-trivial equations are for $i=k$ (equivalently $j=k$), providing a set of $d$ independent equations
\[\pder{f}{q_k}=0.\]

For $a,b>d$ and $c\leq d$, let $i=a-d,j=b-d,k\in\{1,d\}$, we get the following equations:
\[f^{-2}\pder{f}{q_k}L_{ij}+\pder{f}{p_i}\delta_{jk}-\pder{f}{p_j}\delta_{ik}-f\pder{f^{-1}L_{ij}}{q_k}=0.\]
Using the independence of $f$ on $q$s, we obtain
\begin{equation}
\label{d-grad_eq}
    \pder{f}{p_i}\delta_{jk}-\pder{f}{p_j}\delta_{ik}-\pder{L_{ij}}{q_k}=0.
\end{equation}
That contains the gradient equation. There are $d^3$ not-independent equations. If $k\neq i,j$ we obtain $\frac{1}{2}d(d-1)(d-2)$ independent equations
\begin{equation*}
    \pder{L_{ij}}{q_k}=0.
\end{equation*}
Due to antisymmetry of \eqref{d-grad_eq}, $k=i$ gives us the same equations of $k=j$. So, considering $k=j$ and $i\neq j$, we have $d(d-1)$ independent equations
\begin{equation*}
    \pder{f}{p_i}=\pder{L_{ij}}{q_j},
\end{equation*}
which give us constraints on $L_{ij}(q,p)$ and a gradient equation for $f(p)$. The constraints read \[\pder{L_{ij}}{q_j}=\pder{L_{ik}}{q_k},\,\forall i,j,k\leq d,\]
and give us the general form of $L_{ij}$
\begin{equation}
    \label{d-L_form}
    L_{ij}(q,p)=S_{ij}(p)-g_j(p)q_i+g_i(p)q_j,
\end{equation}
where $S_{ij}=-S_{ji}$ and $g_i$ are a set of smooth functions on $p$ only.\\
Thus, the remaining equations are simply recast as the gradient equation
\begin{equation}
\label{fin-grad_eq}
    \pder{f}{p_i}=g_i(p).
\end{equation}
Let us consider this differential equation defined on a simply connected subset $U\subset \R^d$ and define the \emph{1}-form $g=\sum_{i=1}^dg_idp_i$. The equation is equivalent to $df=g$. Hence, necessary and sufficient condition for integrability is $dg=0$.\\
On the other hand, considering a given $f(p)$, the Eq.~\eqref{fin-grad_eq} fully characterizes $L_{ij}$ modulo a function on the momenta without any further request.\\\\
For $a,b,c>d$, a new equation appears, which is not present in the \emph{2}-dimensional case:
\begin{equation}
    \label{d-strange_eq}
    2\left(\pder{f}{p_k}L_{ij}+\pder{f}{p_i}L_{jk}+\pder{f}{p_j}L_{ki}\right)-f\left(\pder{L_{ij}}{p_k}+\pder{L_{jk}}{p_i}+\pder{L_{ki}}{p_j}\right)=0.
\end{equation}
The implication of this ``strange equation" is not clear immediately.\\
We can study it considering the general form of $L_{ij}$ given before, with $S_{ij}\equiv 0$. The first term vanishes 
\begin{align*}
    &\pder{f}{p_k}L_{ij}+\pder{f}{p_i}L_{jk}+\pder{f}{p_j}L_{ki}\\
    &=g_k(-g_jq_i+g_iq_j)+g_i(-g_kq_j+g_jq_k)+g_j(-g_iq_k+g_kq_i)\\
    &=0,
\end{align*}
while the second term contains the integrability conditions
\begin{align*}
   &\pder{L_{ij}}{p_k}+\pder{L_{jk}}{p_i}+\pder{L_{ki}}{p_j}=\\
   &=-\pder{g_j}{p_k}q_i+\pder{g_i}{p_k}q_j-\pder{g_k}{p_i}q_j+\pder{g_j}{p_i}q_k-\pder{g_i}{p_j}q_k+\pder{g_k}{p_j}q_i\\
   &=\left(\pder{g_k}{p_j}-\pder{g_j}{p_k}\right)q_i+\left(\pder{g_i}{p_k}-\pder{g_k}{p_i}\right)q_j+\left(\pder{g_j}{p_i}-\pder{g_i}{p_j}\right)q_k.
\end{align*}
Hence, considering the gradient equation satisfied, the strange equation becomes a constraint on the $S_{ij}$ functions, which are not fixed by the form of $f(p)$:
\begin{equation}
    \label{d_s_eq}
    2\left(\pder{f}{p_k}S_{ij}+\pder{f}{p_i}S_{jk}+\pder{f}{p_j}S_{ki}\right)-f\left(\pder{S_{ij}}{p_k}+\pder{S_{jk}}{p_i}+\pder{S_{ki}}{p_j}\right)=0.
\end{equation}
In reason of the previous discussion, $S_{ij}\equiv 0$ is a solution of the strange equation. 

\subsubsection{The \emph{3}-dimensional case}
\label{3-d}
 The simplest case in which the strange equation appears is the three dimensional one.
 Following the results of the previous Section, in dimension \emph{3} the closure condition implies $\binom{6}{3}=20$ equations. \\
 Ten of them are trivial identities if we consider $f(p)$ depending only on momenta variables. 
 We then proceed to examine the remaining 9+1 equations. Let us consider the first nine of these that correspond to \eqref{d-grad_eq}:
 \begin{equation}
 \label{3-d system}
    \begin{cases}
        (145) & \pder{f}{p_2}+\pder{L_{12}}{q_1}=0,\\
        (146) & \pder{f}{p_3}+\pder{L_{13}}{q_1}=0,\\
        (156) & \pder{L_{23}}{q_1}=0,\\
        (245) & \pder{f}{p_1}-\pder{L_{12}}{q_2}=0,\\
        (246) & \pder{L_{13}}{q_2}=0,\\
        (256) & \pder{f}{p_3}+\pder{L_{23}}{q_2}=0,\\
        (345) & \pder{L_{12}}{q_3}=0,\\
        (346) & \pder{f}{p_1}-\pder{L_{13}}{q_3}=0,\\
        (356) & \pder{f}{p_2}-\pder{L_{23}}{q_3}=0.
    \end{cases}
\end{equation}
As already discussed, that system brings out the following form for $L_{ij}$
\begin{align*}
    L_{12}(q,p)=S_{12}(p)-P(p)q_1+Q(p)q_2,\\
    L_{23}(q,p)=S_{23}(p)-R(p)q_2+P(p)q_3,\\
    L_{13}(q,p)=S_{13}(p)-R(p)q_1+Q(p)q_3.
\end{align*}
Here, $S_{ij},P,Q,R$ are smooth functions of $p_1,p_2,p_3$. In such a way, the system \eqref{3-d system} is recast as three independent equations
\begin{equation}
\label{3-grad_eq}
    \begin{cases}
        \pder{f}{p_1}=Q(p_1,p_2,p_3),\\
        \pder{f}{p_2}=P(p_1,p_2,p_3),\\
        \pder{f}{p_3}=R(p_1,p_2,p_3).
    \end{cases}
\end{equation}
Again we obtain a gradient equation in an open subset $U\subset\R^3$. If $U$ is simply connected, the system is solvable if and only if
\[\pder{P}{p_1}-\pder{Q}{p_2}=0,\ \ \ \pder{R}{p_1}-\pder{Q}{p_3}=0,\ \ \ \pder{P}{p_3}-\pder{R}{p_2}=0,\]
that is the irrationality of the vector field $\mathbf{V}=(Q,P,R)$. In this case, the solution for $f(p)$ is given by fixing a curve $\gamma\subset U$ with endpoint $(p_1,p_2,p_3)$ 
\begin{equation}
    f(p_1,p_2,p_3)=\int_{\gamma}\mathbf{V}\cdot d\gamma.
\end{equation}
In dimension three, the strange equation has only one component, which reads
\[2\left(\pder{f}{p_1}L_{23}+\pder{f}{p_2}L_{31}+\pder{f}{p_3}L_{12}\right)-f\left(\pder{L_{12}}{p_3}+\pder{L_{23}}{p_1}+\pder{L_{31}}{p_2}\right)=0.\]
Considering the gradient equation \eqref{3-grad_eq} satisfied, we obtain a constraint on $S_{ij}$:
\[2\left(Q\,S_{23}+P\,S_{31}+R\,S_{12}\right)-f\left(\pder{S_{12}}{p_3}+\pder{S_{23}}{p_1}+\pder{S_{31}}{p_2}\right)=0.\]
We note that solutions of the following system are solutions of the above equation as well:
\begin{equation*}
    \begin{cases}
        2R\,S_{12}-f\pder{S_{12}}{p_3}=0,\\
        2Q\,S_{23}-f\pder{S_{23}}{p_1}=0,\\
        2P\,S_{31}-f\pder{S_{31}}{p_2}=0.
    \end{cases}
\end{equation*}
In the subset of $\R^3$ where $S_{ij}(p)\neq 0$ and $f(p)\neq 0$, the system can be written compactly as
\begin{equation}
    \frac{1}{S_{ij}}\pder{S_{ij}}{p_k}=2\frac{1}{f}\pder{f}{p_k}\ \ \ \mathrm{with}\  i\neq j\neq k.
\end{equation}
Hence, a possible solution is $S_{ij}=\pm f^2e^{c_{ij}(p_i,p_j)}$, where $c_{ij}=c_{ji}$ are arbitrary functions on $p_i,p_j$ only.
\subsubsection{Interesting example: Kempf-Mangano-Mann 3D-GUP}
\label{L_angular-3}
We now want to analyze the ``angular momentum" case in three dimensions but from the perspective of a fixed $f(p)$.\\
Let $f(p)=\beta(p_1^2+p_2^2+p_3^2)+c$, with $\beta$ strictly positive. This choice forces $Q=2\beta p_1,\,P=2\beta p_2,\,R=2\beta p_3$ because of \eqref{3-grad_eq}. Hence,
\[L_{ij}=S_{ij}(p)+2\beta(q_jp_i-q_ip_j).\]
The Kempf-Mangano-Mann (KMM) GUP theory is recovered for $c=1$ and $S_{ij}\equiv 0$. However, as we already said, this is not the unique choice for $S_{ij}$. Nevertheless, considering $\beta$ the control parameter and requiring that the usual Heisenberg algebra is recovered for $\beta \to 0$, it is possible to exclude some $S_{ij}$ functions. For instance, $S_{ij}= f^2e^{c_{ij}(p_i,p_j)}$ is an admissible solution of \ref{d_s_eq} but it does not have the correct limit because $f\xrightarrow[\beta\to 0]{} c$, forcing $c=1$, and so $S_{ij}\xrightarrow[\beta\to 0]{} e^{c_{ij}(p_i,p_j)}$, which cannot be zero.
\textcolor{blue}For the same reason discussed in Section \ref{l_angular}, here we are not considering a possible dependence of the $c_{ij}$ functions, nor of the constant $c$, on the control parameter $\beta$.

 \section{Maggiore GUP as an application}
 \label{Sec:Maggiore}
 The mathematical relations we have derived in Section \ref{Sec:build_f} are completely general and are based only on the request that the \emph{2}-form $\omega$ \eqref{Sp-form} is symplectic, without further assumptions on the $f$ and $L_{ij}$ functions. \\
 Nevertheless, usually, in constructing a GUP theory other constraints are imposed on the general algebra \eqref{GUP}.
 The most common additional demands include the rotational invariance of the algebra and the condition that the $L_{ij}$ functions satisfy an angular momentum algebra. \\
 This specific subclass of GUP theories is extensively discussed in  \cite{Maggiore:1993kv, Fadel:2021hnx} within the quantum framework, where formal and physical arguments supporting this choice are presented, showing in particular how it still maintains a certain level of generality. \\
 In these works the authors consider the following structure for the GUP algebra:
 \begin{align}
 \label{Maggiore_alg}
    &[\hat p_i,\hat p_j]=0, \nonumber\\
    &[\hat q_i,\hat q_j]=\frac{\hbar^2}{\kappa^2 c^2}a(\hat p)i\epsilon_{ijk}\hat J_{k},\\    
    &[\hat q_i,\hat p_j]= i\hbar f(\hat p)\delta_{ij}. \nonumber
 \end{align}
 where $J_k$ satisfies \[[\hat J_i,\hat q_j]=i\epsilon_{ijk}\hat q_{k},\ \ \ [\hat J_i,\hat p_j]=i\epsilon_{ijk}\hat p_{k}, \ \  \ [\hat J_i,\hat J_j]=i \epsilon_{ijk}\hat{J}_k ,\]
 $a$ and $f$ depend only on the momentum modulus $\rho$, while $\kappa$ is a deformation parameter with the dimension of a mass and $c$ is the speed of light in vacuum.
 
 By means of Jacobi identities, they are able to find two equations relating $f$ and $a$:
 \begin{align}
 \label{Maggiore_conditions1}
    & \frac{d a(\hat \rho)}{d \rho} \ \hat{\mathbf{p}} \cdot \hat{\mathbf{J}}=0,  \\
  \label{Maggiore_conditions2}
    & f(\hat \rho) \frac{d f(\hat \rho)}{d \rho}=-\frac{a(\hat \rho) \hat \rho}{\kappa^2 c^2}.
 \end{align}
 Two cases can be distinguished: 
 \begin{itemize}
     \item $\hat{\mathbf{p}} \cdot \hat{\mathbf{J}}=0$.
     This case arises  when $\hat{\mathbf{J}}$ is regarded only as orbital angular momentum. 
     Due to the zero scalar product the constrains \eqref{Maggiore_conditions1} is removed, leaving the choice of $a$ arbitrary.
     Fixed a suitable $a$ it is possible to determine correctly $f$ or viceversa.
     \item $\hat{\mathbf{p}} \cdot \hat{\mathbf{J}}\neq 0$.
     This is the most general situation, happening when we are considering $\hat{\mathbf{J}}$ as a total angular momentum, including spin. \\
     In this case \eqref{Maggiore_conditions1} sets $a(\hat \rho)=\pm k$, where $k$ is a constant. By rescaling, we can fix $a(\hat p)=\pm 1$ and from \eqref{Maggiore_conditions2} we obtain, modulo a sign, a unique solution for $f$, namely $f(\hat \rho)=\sqrt{1 \mp \frac{\hat{\rho}^2}{\kappa^2 c^2}}$. \\
     Once again we have set the integration constant equal to one, for the reason discussed above.
 \end{itemize}

 The final algebra then reads:
 \begin{align*}
    &[\hat q_i,\hat q_j]=\mp \frac{\hbar^2}{\kappa^2}i\epsilon_{ijk}\hat J_{k},\\    
    &[\hat q_i,\hat p_j]= i\hbar \delta_{ij} \sqrt{1 \pm \tfrac{\hat \rho^2}{\kappa^2 c^2}}.
 \end{align*}
 As it should be clear from the derivation, these specific GUP models acquire relevance since they are the most general models, defined only by the imposition of Jacobi identities and the inclusion of the spin operator.
 We stress that the generality of the model has to be intended in the sense specified above.

 In light of this, in this section our aim is to discuss this particular scheme at the classical level, further constrain the general relations for $f$ and $L_{ij}$ we have obtained only asking for the $\omega$ to be symplectic.
 Exactly  as in  \cite{Maggiore:1993kv, Fadel:2021hnx} we limit our discussion to the \emph{3}-dimensional case.

 \subsection{Closure condition and Maggiore algebra}

 Consider the corresponding classical interpretation of the algebra \eqref{Maggiore_alg}:
    \begin{align} 
    \label{Maggiore GUP_classical}
    &\{p_i,p_j\}=0, \nonumber \\
    &\{q_i,q_j\}= a(\rho)\epsilon_{ijk}J_{k},\\  &\{q_i,p_j\}= f(\rho)\delta_{ij}. \nonumber
    \end{align}
 Note that now $J_i$ is a dimension-full quantity and we have absorbed the term $(\kappa c)^{-2}$ in the definition of $a$.\\
 We demand that $f$ and $a$ depend only on the modulus of momentum $\rho$ and that the $J_i$s satisfy the algebra of an angular momentum, that is $\{J_i, q_j\}=\epsilon_{ijk} q_{k},\,\{J_i,p_j\}=\epsilon_{ijk}p_{k}, \{J_i, J_j\}=\epsilon_{ijk} J_{k}$.\\\\
 With specific reference to the Poisson brackets above, the previously derived system of equations \eqref{3-d system} coming from the closure condition, yields the following relations:
    \begin{equation}
    \label{3d_system_Maggiore}
    \begin{cases}
        (145) & \pder{f}{p_2}+a(p)\pder{J_3}{q_1}=0,\\
        (146) & \pder{f}{p_3}-a(p)\pder{J_2}{q_1}=0,\\
        (156) & \pder{J_1}{q_1}=0,\\
        (245) & \pder{f}{p_1}-a(p)\pder{J_3}{q_2}=0,\\
        (246) & \pder{J_2}{q_2}=0,\\
        (256) & \pder{f}{p_3}+a(p)\pder{J_1}{q_2}=0,\\
        (345) & \pder{J_3}{q_3}=0,\\
        (346) & \pder{f}{p_1}+a(p)\pder{J_2}{q_3}=0,\\
        (356) & \pder{f}{p_2}-a(p)\pder{J_1}{q_3}=0.
    \end{cases}
    \end{equation}
By the same considerations developed for the system \eqref{3-d system} we can obtain the following general structure for the $J$s functions:
    \begin{align} \label{J_functions}
    &J_1(q,p)=s_1(p)-R(p)q_2+P(p)q_3, \nonumber\\
    &J_2(q,p)=s_2(p)+R(p)q_1-Q(p)q_3, \\
    &J_3(q,p)=s_3(p)-P(p)q_1+Q(p)q_2.  \nonumber
    \end{align}
By imposing an additional structure, namely the angular momentum algebra to the $J$s functions, we expect to be able to further constrain the previous expressions and relations.\\
As a first thing, we are going to explicitly evaluate the Poisson brackets $\{J_m,p_l\}$. This procedure leads the following result:
    \begin{equation}
    \{J_m,p_l\}=\pder{J_m}{q_l}f(p):=\epsilon_{mlk}p_k.
    \end{equation}
 From here, using the $J$s formulae \eqref{J_functions}, we consistently find:
    \begin{equation}
    R(p)=-f^{-1}p_3, \quad P(p)=-f^{-1}p_2, \quad Q(p)=-f^{-1}p_1.
    \end{equation}
This allows us to rewrite the functions \eqref{J_functions} as:
    \begin{align} 
    \label{J_functions_2}
    &J_1(q,p)=s_1(p)+f^{-1}(p)(p_3q_2 -p_2q_3),\nonumber \\
    &J_2(q,p)=s_2(p)+f^{-1}(p)(p_1q_3-p_3q_1), \\
    &J_3(q,p)=s_3(p)+f^{-1}(p)(p_2q_1-p_1q_2).  \nonumber
    \end{align}
Accordingly, the gradient equation for $f$ that can be extracted from the system \eqref{3d_system_Maggiore} becomes:
    \begin{equation} \label{f_system}
    \begin{cases}
        &\pder{f}{p_1}+a(p)p_1 f^{-1}(p)=0, \\
        &\pder{f}{p_2}+a(p)p_2 f^{-1}(p)=0,  \\
        &\pder{f}{p_3}+a(p)p_3 f^{-1}(p)=0.
        \end{cases}
    \end{equation}
We now proceed in calculating explicitly the set of Poisson brackets $ \{J_m,q_l\}$, which lead to the following relation:
    \begin{equation}
     \{J_m,q_l\}=-\pder{J_m}{p_l}f(p)+a(p) \pder{J_m}{q_i}\epsilon_{ilk}J_{k}:=\epsilon_{mlk}q_k.
    \end{equation}
 By means of the expressions \eqref{J_functions_2} for the $J$s functions we obtain a system of equations, which we can compactly write as:
 \begin{equation*}
     -\pder{s_m}{p_l}f+\frac{a}{f} (s_mp_l-s_ip_j\delta_{ij}\delta_{ml})+\epsilon_{ijm}q_ip_j\left(\frac{1}{f}\pder{f}{p_l}+\frac{a}{f^2}p_l\right)=0.
 \end{equation*}

 This set of equations can be simplified by considering the system \eqref{f_system}, leading to the following partial differential equations system:
    \begin{equation} \label{qPB_system}
        \begin{cases}
            & f^2(p)\pder{s_1}{p_1}+a(p)(s_2(p)p_2+s_3(p)p_3)=0,\\
            & f^2(p)\pder{s_2}{p_2}+a(p)(s_1(p)p_1+s_3(p)p_3)=0, \\
            & f^2(p)\pder{s_3}{p_3}+a(p)(s_1(p)p_1+s_2(p)p_2)=0, \\
            & f^2(p)\pder{s_1}{p_2}-a(p)p_2s_1(p)=0, \\
            & f^2(p)\pder{s_1}{p_3}-a(p)p_3s_1(p)=0, \\
            & f^2(p)\pder{s_2}{p_1}-a(p)p_1s_2(p)=0, \\
            & f^2(p)\pder{s_2}{p_3}-a(p)p_3s_2(p)=0, \\
            & f^2(p)\pder{s_3}{p_1}-a(p)p_1s_3(p)=0, \\
            & f^2(p)\pder{s_3}{p_2}-a(p)p_2s_3(p)=0.
        \end{cases}
    \end{equation}

 Finally we carry out the calculations regarding the last set of Poisson brackets for the angular momentum algebra:
    \begin{equation}
         \{J_m,J_n\}=\left(\pder{J_m}{q_i}\pder{J_n}{p_j}-\pder{J_m}{p_i}\pder{J_n}{q_j}\right)\delta_{ij}f+\pder{J_m}{q_i}\pder{J_n}{q_j}a \epsilon_{ijk}J_k:= \epsilon_{mnl}J_l.
    \end{equation}
 Due to the antisymmetry of the brackets, we need to evaluate just three of them:
   \begin{align*} \label{s_system}
       \begin{split}
       \{J_1,J_2\}=&f^{-2}(p)a(p)p_3 \mathbf{p}\cdot \mathbf{J}+J_3-s_3(p)+p_3\left(\pder{s_1}{p_1}+\pder{s_2}{p_2}\right)\\
       &-p_2\pder{s_2}{p_3}-p_1\pder{s_1}{p_3}:=J_3,
       \end{split}
       \\
        \begin{split}
       \{J_2,J_3\}=&f^{-2}(p)a(p)p_1 \mathbf{p}\cdot \mathbf{J}+J_1-s_1(p)+p_1\left(\pder{s_2}{p_2}+\pder{s_3}{p_3}\right)\\
       &-p_3\pder{s_3}{p_1}-p_2\pder{s_2}{p_31}:=J_1,
       \end{split}
       \\
        \begin{split}
       \{J_3,J_1\}=&f^{-2}(p)a(p)p_2 \mathbf{p}\cdot \mathbf{J}+J_2-s_2(p)+p_2\left(\pder{s_1}{p_1}+\pder{s_3}{p_3}\right)\\
       &-p_3\pder{s_3}{p_2}-p_1\pder{s_1}{p_2}:=J_2.
       \end{split}
       \\
   \end{align*}
 Taking now into account the relations established by the system \eqref{qPB_system} and the derived form \eqref{J_functions_2} of the $J$s functions, we can rewrite these conditions on the Poisson brackets as a simple algebraic system for the $s_{i}$:
    \begin{equation}
    \label{S=0-system}
        \begin{cases}
            &f^{-2}a p_1 \mathbf{p}\cdot \mathbf{s}+s_1=0, \\
            &f^{-2}a p_2 \mathbf{p}\cdot \mathbf{s}+s_2=0,\\
            &f^{-2}a p_3 \mathbf{p}\cdot \mathbf{s}+s_3=0.
        \end{cases}
    \end{equation}
 As a homogeneous linear system, its determinant is $f^{-2}(f^2+a(p_1^2+p_2^2+p_3^2))$. The singular points are represented by the equation:
 \begin{equation}
 \label{det_f}
     a(p)(p_1^2+p_2^2+p_3^2)=-f^2(p).
 \end{equation}
 Clearly, this equation admits solution only for $a(p)< 0$. 
 If it is satisfied in an open subset of $\R^3$, then we can put the $a(p)$ derived from the above equation in the gradient equation \eqref{3-grad_eq}.
 As a result, we obtain a unique solution for $f$, which is $f(p)=\kappa \sqrt{p_1^2+p_2^2+p_3^2}$.
 This solution is incompatible with the undeformed limit, since there is no any possible dependence of $\kappa$ on the deformation parameter $\beta$ which results in $f\xrightarrow[\beta\to 0]{} 1$. \\
 As a consequence, the equation \eqref{det_f} can be satisfied only on a submanifold with at least codimension one.
 Where in $\R^3$ the determinant is not singular, the unique solution of \eqref{S=0-system} is $\mathbf{s}=0$. Hence, since $s_i$ is a smooth function everywhere null but on a closed surface, for continuity it must be zero everywhere in $\R^3$.\\
 It is trivial to verify that this solution is compatible with the system \eqref{qPB_system}.\\\\
 Now that we have further specified the form of the $J$s functions, we analyze the remaining condition to be satisfied in order to have the closure of the \emph{2}-form $\omega$.
 This is exactly the Eq.~\eqref{d-strange_eq}, the specific form of which in this context is:
    \begin{equation}
    \label{Maggiore_strange_equation}
    2 a^2 f^{-1}\left(J_1 p_1+J_2 p_2+J_3 p_3\right)+f\left(\pder{(a J_1)}{p_1}+\pder{(a J_2)}{p_2}+\pder{(a J_3)}{p_3}\right)=0.
    \end{equation}
 By employing the last obtained form of the $J$s we get:
   \begin{equation}
   \label{Maggiore_preeq}
   J_1\pder{a}{p_1}+J_2\pder{a}{p_2}+J_3\pder{a}{p_3}=0.
   \end{equation}
 If we now recall our assumption of $a$ depending only on the modulus $\rho=\sqrt{p_1^2+p_2^2+p_3^2}$, we finally obtain the equation:
    \begin{equation} \label{Maggiore_eq}
        \mathbf{p}\cdot\mathbf{J}\; \pder{a}{\rho}=0,
    \end{equation}
 which is the same of \eqref{Maggiore_conditions1}. \\
 Nevertheless, due the explicit form of the $J$s and specifically to the fact that the $s_i$ functions have to be zero, it is immediate to show that $\mathbf{p}\cdot\mathbf{J}$ is always equals to zero in this context.
 This fact leads us to conclude that \eqref{Maggiore_eq} is always satisfied for any smooth, rotational invariant $a(\rho)$ and rules out the case in which $\mathbf{p}\cdot\mathbf{J} \neq 0$. 
 From here we can also deduce the potential role played by the functions $s_i$, as the only part of the $J$s able to ensure the condition  $\mathbf{p}\cdot\mathbf{J} \neq 0$.
 As a consequence, we conclude that the generality of the solution $f(p)=\sqrt{1\mp p^2}$ in the quantum framework, seems to be lost at the classical level, being just one of the possible GUP models.

 Finally we can write down the general solution for $f$ by examining the gradient equation \eqref{f_system}.
 Being $a(\rho)$ rotational invariant and considering that $\rho$ is the length $\rho=\sqrt{p_1^2+p_2^2+p_3^2}$, the system \eqref{f_system} can be recast in a gradient equation for $g=f^2$ in $\mathbf{p}=(p_1,p_2,p_3)\in \R^3$, completely equivalent to \eqref{Maggiore_conditions2}:
 \begin{align} \label{f^2_system}
    \nabla_{\mathbf{p}} g=-2a(\rho)\mathbf{p}.
    \end{align}
 Clearly, the right hand side is irrotational
\[\nabla_{\mathbf{p}}\times(a(\rho)\mathbf{p})=a(\rho)\nabla_{\mathbf{p}}\times\mathbf{p}+\nabla_{\mathbf{p}} a\times\mathbf{p}=\frac{1}{\rho}\pder{a}{\rho}\mathbf{p}\times\mathbf{p}=0.\]
 Then, if $a(\rho)$ is defined on a simply connected domain, the system admits a solution:
 \begin{equation}
    \label{Maggiore_f}
    f(p)=\left(-2 \int_{\gamma} a(p')\mathbf{p}'\cdot d\mathbf{p}'\right)^{\frac{1}{2}}=\left(-2 \int_{0}^{\sqrt{p_1^2+p_2^2+p_3^2}} a(\rho)\rho d\rho+c\right)^{\frac{1}{2}}.
\end{equation}

\subsection{Kempf-Mangano-Mann case in Maggiore's scheme}
One of the most studied GUP algebra is the one first presented in \cite{Kempf:1994su}.\\ 
In Section~\ref{l_angular} we have already shown how this particular algebra can be recovered by choosing the $L_{ij}$ functions as proportional to the ordinary angular momentum and in Section~\ref{L_angular-3} we were able to derive its algebra by fixing the $f$ function and discussing the limit $\beta \to 0$.
Precisely, the full agreement is obtained by setting $S_{ij}\equiv0$ and the integration constant $c$ equal to one.\\\\
Here we briefly show how the KMM algebra can be reproduced  in Maggiore's scheme as well, unveiling the presence of the rotation generators in the modified Poisson brackets of the configuration variables.\\
Starting from the gradient equation \eqref{f_system} in the form:
\begin{equation}
    \nabla_{\mathbf{p}} f +f^{-1} a(\rho) \mathbf{p}=0.
\end{equation}
We can fix the desired $f$ and solve algebrically the system for $a$.
Hence, setting $f(\rho)=1+ \beta \rho^2$, we easily obtain $a(\rho)=-2 \beta (1+ \beta \rho^2) $.
According to the algebra \eqref{Maggiore GUP_classical}, we should have:
\begin{equation}
    \{q_i, q_j\}=a(\rho) \epsilon_{ijk} J_k =-2\beta (1+ \beta \rho^2) \epsilon_{ijk} J_k.
\end{equation}
At the same time we know the exact form of the $J_{k}$ functions from \eqref{J_functions_2}:
\begin{equation}
    J_{k}=\frac{1}{2}f^{-1}\epsilon_{ijk}(q_i p_j-q_j p_i)=\frac{1}{1+\beta \rho^2}\epsilon_{ijk}q_i p_j.
\end{equation}
Putting everything together we finally obtain:
\begin{equation}
    \{q_i, q_j\}=-2 \beta (q_i p_j-q_j p_i),
\end{equation}
which is the correct expression of the KMM algebra at the classical level and the same formula obtained in Section~\ref{L_angular-3}, with the identification previously mentioned, i.e. $S_{ij}=0$ and $c=1$.

\section{Verifying the Conjecture}
\label{Sec:Proof}
In this Section, we are going to provide some cases that satisfy the Conjecture \ref{Theo}. Let us start from the following Proposition (Prop.8.11 in \cite{Libermann_1987}):
\begin{proposition}
    \label{Prop_equiv}
    A symplectic form on $\M$ is equivalent to a non-degenerate Poisson structure.
\end{proposition}
The equivalence is given identifying the Poisson bivector field $\pi$ \cite{Weinstein_1998} with the inverse of the symplectic form $\omega$, i.e., in coordinates, $\pi^{ab}=\omega^{ab}$.\\\\
Consider the fundamental Poisson brackets expressed in local coordinates $x^a=(q_1,\dots,q_d, p_1,\dots,p_d)$ as in \eqref{Poisson} , where $L_{ij}$ and $f$ are smooth functions of the coordinates, and $f$ is strictly positive on $\M$. If these Poisson brackets satisfy the Jacobi identities, then they define a non-degenerate Poisson structure. The non-degeneracy can be easily checked considering the bivector field in coordinates
\begin{equation}
   \pi^{ab}:=\{x^a,x^b\}=
   \begin{pmatrix}
    L & f\mathrm{id}_d\\
    -f\mathrm{id}_d & 0
   \end{pmatrix},
 \end{equation}
for which $\det (\pi^{ab})=f^{2d}>0$. From this and Prop.~\ref{Prop_equiv}, the following Lemma holds:
\begin{lemma}
    The \emph{2}-form $\omega$ defined in \eqref{Sp-form} is a symplectic form if and only if the Poisson brackets in \eqref{Poisson} satisfy the Jacobi identity.
\end{lemma}

Since a rigorous and consistent quantization procedure does not exist for GUP theories, we just check that the quantum Jacobi identity is satisfied if we consider $f$ and $L$ as in Section~\ref{d-d} for two possible 'natural' operator ordering in the quantum theory, namely all the momenta to the left and all the momenta to the right. \\ 

Let us consider a GUP theory in the class \eqref{GUP}, with $L$ and $f$ that, as functions on the phase space, satisfy \eqref{d-q_indip},\eqref{d-L_form},\eqref{fin-grad_eq}. The main problem is the choice of the operator ordering in $L_{ij}(\q,\p)$, since, in the classical functions, monomials of the kind $p_iq_i$ appear. The other functions, which depend only on $p$, do not have any ambiguity in the ordering.\\
Let us first check the Jacobi identity for the fundamental Poisson brackets that do not depend on $L_{ij}$. On the momenta, the commutation is trivially null $[\p_i,[\p_j,\p_k]]+[\p_j,[\p_k,\p_i]]+[\p_k,[\p_i,\p_j]]=0$.\\
Moreover, we have
\begin{align*}
    &[\p_i,[\p_j,\q_k]]+[\p_j,[\q_k,\p_i]]+[\q_k,[\p_i,\p_j]]\\
    &=-i\hbar\delta_{jk}[\p_i,f(\p)]+i\hbar\delta_{ik}[\p_j,f(\p)]=0.
\end{align*}
Notice that, independently from the chosen ordering for $L_{ij}(\q,\p)$, we get 
\begin{equation}
    [\p_k,L_{ij}(\q,\p)]=i\hbar\left(\delta_{ik}g_j(\p)f(\p)-\delta_{jk}g_i(\p)f(\p)\right).
\end{equation}
Furthermore, since there are no ambiguities in $f(p)$, its derivative is well-defined $\pder{f}{p_i}(\p)$ in its clear meaning and so it is not difficult to prove that:
\begin{equation}
    [\q_k,f(\p)]=i\hbar f(\p)\pder{f}{p_k}(\p)=:i\hbar f(\p)g_k(\p).
\end{equation}
This leads us to conclude that another commutator is free from ambiguities:
\begin{align*}
    &[\q_i,[\q_j,\p_k]]+[\q_j,[\p_k,\q_i]]+[\p_k,[\q_i,\q_j]]\\
    &=i\hbar\delta_{jk}[\q_i,f(\p)]-i\hbar\delta_{ik}[\q_j,f(\p)]+i\hbar[\p_k,L_{ij}(\q,\p)]\\
    &=-\hbar^2\left(\delta_{jk}f(\p)g_i(\p)-\delta_{ik}f(\p)g_j(\p)+\delta_{ik}g_j(\p)f(\p)-\delta_{jk}g_i(\p)f(\p)\right)\\
    &=0.
\end{align*}

We now need to fix an operator ordering in the theory to proceed. Let us first consider all the momenta on the left, so the commutation between configuration coordinates reads
\begin{equation}
    [\q_i,\q_j]=i\hbar \left(S_{ij}(\p)-g_j(\p)\q_i+g_i(\p)\q_j\right),
\end{equation}
from which, we can write the double commutator:
\begin{align*}
    &[\q_k,[\q_i,\q_j]]\\
    &=i\hbar \Big([\q_k,S_{ij}(\p)]-[\q_k,g_j(\p)]\q_i-g_j(\p)[\q_k,\q_i]+[\q_k,g_i(\p)]\q_j+g_i(\p)[\q_k,\q_j]\Big)\\
    &=-\hbar^2 \Bigg(f(\p)\pder{S_{ij}}{p_k}(\p)-f(\p)\pder{g_j}{p_k}(\p)\q_i-g_j(\p)L_{ki}+f(\p)\pder{g_i}{p_k}(\p)\q_j+g_i(\p)L_{kj}\Bigg).\\
\end{align*}
Fixing $S_{ij}\equiv 0$, the last Jacobi identity is satisfied
\begin{align*}
    &\frac{1}{\hbar^2}\left([\q_i,[\q_j,\q_k]]+[\q_j,[\q_k,\q_i]]+[\q_k,[\q_i,\q_j]]\right)\\
    &=-f(\p)\pder{g_k}{p_i}(\p)\q_j-g_k(\p)L_{ij}+f(\p)\pder{g_j}{p_i}(\p)\q_k+g_j(\p)L_{ik}\\
    &\ \ \ \ -f(\p)\pder{g_i}{p_j}(\p)\q_k-g_i(\p)L_{jk}+f(\p)\pder{g_k}{p_j}(\p)\q_i+g_k(\p)L_{ji}\\
    &\ \ \ \ -f(\p)\pder{g_j}{p_k}(\p)\q_i-g_j(\p)L_{ki}+f(\p)\pder{g_i}{p_k}(\p)\q_j+g_i(\p)L_{kj}\\
    &=f(\p)\left(\pder{g_k}{p_j}(\p)-\pder{g_j}{p_k}(\p)\right)\q_i+f(\p)\left(\pder{g_i}{p_k}(\p)-\pder{g_k}{p_i}(\p)\right)\q_j\\
    &\ \ \ \ +f(\p)\left(\pder{g_j}{p_i}(\p)-\pder{g_i}{p_j}(\p)\right)\q_k-2g_k(\p)\left(-g_j(\p)\q_i+g_i(\p)\q_j\right)\\
    &\ \ \ \ -2g_i(\p)\left(-g_k(\p)\q_j+g_j(\p)\q_k\right)-2g_j(\p)\left(-g_i(\p)\q_k+g_k(\p)\q_i\right)\\
    &=0.
\end{align*}
For a not identically null $S_{ij}$, the Jacobi identity imposes
\begin{align*}
    &0=\frac{1}{\hbar^2}\left([\q_i,[\q_j,\q_k]]+[\q_j,[\q_k,\q_i]]+[\q_k,[\q_i,\q_j]]\right)\\
    &=f(\p)\pder{S_{jk}}{p_i}(\p)-g_k(\p)S_{ij}(\p)+g_j(\p)S_{ik}(\p)+f(\p)\pder{S_{ki}}{p_j}(\p)\\
    &\ \ \ \ -g_i(\p)S_{jk}(\p)+g_k(\p)S_{ji}(\p)+f(\p)\pder{S_{ij}}{p_k}(\p)-g_j(\p)S_{ki}(\p)+g_i(\p)S_{kj}(\p)\\
    &=f(\p)\left(\pder{S_{ij}}{p_k}(\p)+\pder{S_{jk}}{p_i}(\p)+\pder{S_{ki}}{p_j}(\p)\right)\\
    &\ \ \ \ -2\left(g_k(\p)S_{ij}(\p)+g_i(\p)S_{jk}(\p)+g_j(\p)S_{ki}(\p)\right),
\end{align*}
which is the strange equation \eqref{d_s_eq}. If $S_{ij}$ fulfill the classic equation the Jacobi identity is verified and no ambiguities in the ordering arise in the final expression, where everything is just a function of the momentum operator.\\
Nevertheless, the choice of the ordering of the $L_{ij}$ operators is relevant for this last Jacobi identity, even after $S_{ij}$ is fixed by solving the strange equation. 
That any ordering might lead to the same result and thus verify the validity of the latter Jacobi identity is neither clear nor straightforward to confirm in the general case.
However, we can notice that in the other possible common choice with all the momenta on the right, the Jacobi identity is still achieved.\\\\
The converse, namely if the quantum commutators of a GUP in the class \eqref{GUP} satisfy the Jacobi identities then the Poisson brackets of its classical interpretation satisfy them too, is trivial.

\section{Conclusions}
\label{Sec:Concl}
The focus of this work is to explore whether a GUP theory can be consistently and rigorously implemented at the classical level. Assuming the classical system related to GUP theory is defined by a Poisson structure inherited from deformed quantum commutators, we established the corresponding \emph{2}-form, ensuring it is symplectic, i.e., non-degenerate and closed. These conditions lead to a set of equations and relationships that the functions controlling the deformation of the usual Heisenberg algebra must satisfy.

By carefully examining the implications of these equations, we defined a general form for the functions \(L_{ij}\) that introduce non-commutativity in the considered GUP models, along with a gradient equation for the function \(f\) that regulates the deformation of the Poisson brackets between conjugate variables.

When the theory respects these relations, its structure is guaranteed to be symplectic. This ensures that the classical Hamiltonian formulation for such a theory is well-defined and can be effectively used to study the dynamics of systems in the deformed phase space characterized by \eqref{Poisson}. Consistent with symplectic geometry, requiring the \emph{2}-form to be symplectic is equivalent to ask for the validity of the Jacobi identities for the Poisson brackets, a condition explicitly verified in the appendix as a further confirmation.

In the second part of the paper, we examined the consequences of adding another structure to the deformed Poisson brackets. Specifically, following the procedure used at the quantum level in \cite{Maggiore:1993kv, Fadel:2021hnx}, we required that the functions \(L_{ij}\) governing non-commutativity in configuration space satisfy the angular momentum algebra. This specification allowed us to determine the form of these functions, which are the generators of rotations in these GUP theories. The resulting explicit form for these generators excludes the possibility that the scalar product \(p \cdot J\) is nonzero. This is in contrast with the findings reported in \cite{Maggiore:1993kv, Fadel:2021hnx} at the quantum level, where the scenario in which \(p \cdot J \neq 0\) is possible and it is used to extract a GUP model that is, in some ways, more general, accounting for the presence of spin operator. At the classical level, the algebra requirements rule out this possibility, as \(p \cdot J = 0\) within this class of GUP theories, leading to an infinite class of models, formally on the same level, without selecting a specific one.

An interesting follow-up to the analysis we carried out regarding the classical symplectic structure of these theories could be represented by the investigation of relativistic non-commutative models. In this respect, the well-known Snyder space-time \cite{PhysRev.71.38} is a notable representative that will deserve examination within our framework.

Finally, we concluded the article by discussing a Conjecture in order to bridge the classical theory thus formulated and the quantum theory: a properly defined GUP theory can be correctly classically interpreted if and only if the corresponding quantum commutators satisfy the Jacobi identities. While transitioning from quantum to classical theory is straightforward concerning the Jacobi identities for brackets and commutators, the reverse is not obvious due to potential operator ordering ambiguities. If all classical relations ensuring a symplectic structure in GUP theory are assumed valid at the quantum level, this will ensure that the Jacobi identities for all commutators without ordering ambiguities are automatically satisfied. For the remaining commutators, specifically those involving only the generalized coordinates \(q\)s, verifying the Jacobi identity requires fixing an ordering. Although proving that any ordering satisfies the Jacobi identities is generally not possible, the two most natural orderings do, providing potentially a guideline for constructing the quantum theory.

\appendix
\section{Poisson brackets and Jacobi identity}
\label{appendix}
The link between symplectic structure and Poisson structure is well-known. In particular, considering a non-degenerate Poisson structure $\{\cdot,\cdot\}$ on $\M$ and the induced (or induced by) symplectic form $\omega$, the Jacobi identity for $\{\cdot,\cdot\}$ is equivalent to the closure of $\omega$. Hence, the result of Section~\ref{Sec:build_f} can be equivalently derived imposing the Jacobi identity on the fundamental Poisson brackets. In coordinates, $\{x^a,\{x^b,x^c\}\}+\{x^b,\{x^c,x^a\}\}+\{x^c,\{x^a,x^b\}\}=0$.\\\\
Let us considering the fundamental Poisson brackets in \eqref{Poisson}.\\
For $a,b,c>d$, the equations are just trivial identities.\\\\
For $c\leq d,\,a,b>d$, we recover the $q$-independence of $f$ as in \eqref{d-q_indip}
\begin{align*}
    &\{p_i,\{p_j,q_k\}\}+\{p_j,\{q_k,p_i\}\}+\{q_k,\{p_i,p_j\}\}\\
    &=-\{p_i,f\}\delta_{jk}+\{p_j,f\}\delta_{ki}\\
    &=\pder{f}{q_i}\delta_{jk}-\pder{f}{q_j}\delta_{ik}=0.
\end{align*}
For $a,b\leq d,\,c>d$, we obtain the equation \eqref{d-grad_eq}
\begin{align*}
    &\{q_i,\{q_j,p_k\}\}+\{q_j,\{p_k,q_i\}\}+\{p_k,\{q_i,q_j\}\}\\
    &=\{q_i,f\}\delta_{jk}-\{q_j,f\}\delta_{ik}+\{p_k,L_{ij}\}\\
    &=f\pder{f}{p_i}\delta_{jk}-f\pder{f}{p_j}\delta_{ik}-f\pder{L_{ij}}{q_k}=0.
\end{align*}
For $a,b,c\leq d$, the strange equation appears:
\begin{align*}
    &\{q_i,\{q_j,q_k\}\}+\{q_j,\{q_k,q_i\}\}+\{q_k,\{q_i,q_j\}\}\\
    &=\{q_i,L_{jk}\}+\{q_j,L_{ki}\}+\{q_k,L_{ij}\}\\
    &=f\pder{L_{jk}}{p_i}+f\pder{L_{ki}}{p_j}+f\pder{L_{ij}}{p_k}+\sum_{r=1}^d\left(\pder{L_{jk}}{q_r}L_{ir}+\pder{L_{ki}}{q_r}L_{jr}+\pder{L_{ij}}{p_r}L_{kr}\right)\\
    &=0.
\end{align*}
The usual form of the strange equation \eqref{d-strange_eq} is recovered using the gradient equation. Indeed, the second term reads as:
\begin{align*}
    &\sum_{r=1}^d\left(\pder{L_{jk}}{q_r}L_{ir}+\pder{L_{ki}}{q_r}L_{jr}+\pder{L_{ij}}{p_r}L_{kr}\right)\\
    &=\sum_{r=1}^d\Bigg((\pder{f}{p_j}\delta_{kr}-\pder{f}{p_k}\delta_{jr})L_{ir}+(\pder{f}{p_k}\delta_{ir}-\pder{f}{p_i}\delta_{kr})L_{jr}\\
    &\ \ \ \ \ \ \ \ \ \ \ \ \ +(\pder{f}{p_i}\delta_{jr}-\pder{f}{p_j}\delta_{ir})L_{kr}\Bigg)\\
    &=-2\left(\pder{f}{p_k}L_{ij}+\pder{f}{p_i}L_{jk}+\pder{f}{p_j}L_{ki}\right).
\end{align*}

Thus, from these simple computations, we evince the expected equivalence between the two approaches.

\bibliographystyle{elsarticle-num} 
\bibliography{updated_biblio}

\begin{thebibliography}{10}
\expandafter\ifx\csname url\endcsname\relax
  \def\url#1{\texttt{#1}}\fi
\expandafter\ifx\csname urlprefix\endcsname\relax\def\urlprefix{URL }\fi
\expandafter\ifx\csname href\endcsname\relax
  \def\href#1#2{#2} \def\path#1{#1}\fi

\bibitem{Amati:1988tn}
D.~Amati, M.~Ciafaloni, G.~Veneziano, Can space-time be probed below the string size?, Phys. Lett. B 216 (1989) 41--47.
\newblock \href {https://doi.org/10.1016/0370-2693(89)91366-X} {\path{doi:10.1016/0370-2693(89)91366-X}}.

\bibitem{Konishi:1989wk}
K.~Konishi, G.~Paffuti, P.~Provero, Minimum physical length and the generalized uncertainty principle in string theory, Phys. Lett. B 234 (1990) 276--284.
\newblock \href {https://doi.org/10.1016/0370-2693(90)91927-4} {\path{doi:10.1016/0370-2693(90)91927-4}}.

\bibitem{Maggiore:1993rv}
M.~Maggiore, {A Generalized uncertainty principle in quantum gravity}, Phys. Lett. B 304 (1993) 65--69.
\newblock \href {http://arxiv.org/abs/hep-th/9301067} {\path{arXiv:hep-th/9301067}}, \href {https://doi.org/10.1016/0370-2693(93)91401-8} {\path{doi:10.1016/0370-2693(93)91401-8}}.

\bibitem{Kempf:1994su}
A.~Kempf, G.~Mangano, R.~B. Mann, {Hilbert space representation of the minimal length uncertainty relation}, Phys. Rev. D 52 (1995) 1108--1118.
\newblock \href {http://arxiv.org/abs/hep-th/9412167} {\path{arXiv:hep-th/9412167}}, \href {https://doi.org/10.1103/PhysRevD.52.1108} {\path{doi:10.1103/PhysRevD.52.1108}}.

\bibitem{Kempf:1996nk}
A.~Kempf, G.~Mangano, {Minimal length uncertainty relation and ultraviolet regularization}, Phys. Rev. D 55 (1997) 7909--7920.
\newblock \href {http://arxiv.org/abs/hep-th/9612084} {\path{arXiv:hep-th/9612084}}, \href {https://doi.org/10.1103/PhysRevD.55.7909} {\path{doi:10.1103/PhysRevD.55.7909}}.

\bibitem{Segreto:2022clx}
S.~Segreto, G.~Montani, {Extended GUP formulation and the role of momentum cut-off}, Eur. Phys. J. C 83~(5) (2023) 385.
\newblock \href {http://arxiv.org/abs/2208.03101} {\path{arXiv:2208.03101}}, \href {https://doi.org/10.1140/epjc/s10052-023-11480-4} {\path{doi:10.1140/epjc/s10052-023-11480-4}}.

\bibitem{Segreto:2024vtu}
S.~Segreto, G.~Montani, {n-Dimensional non-commutative GUP quantization and application to the Bianchi I model}, Eur. Phys. J. C 84~(8) (2024) 796.
\newblock \href {http://arxiv.org/abs/2401.17113} {\path{arXiv:2401.17113}}, \href {https://doi.org/10.1140/epjc/s10052-024-13145-2} {\path{doi:10.1140/epjc/s10052-024-13145-2}}.

\bibitem{Bosso:2023aht}
P.~Bosso, G.~G. Luciano, L.~Petruzziello, F.~Wagner, {30 years in: Quo vadis generalized uncertainty principle?}, Class. Quant. Grav. 40~(19) (2023) 195014.
\newblock \href {http://arxiv.org/abs/2305.16193} {\path{arXiv:2305.16193}}, \href {https://doi.org/10.1088/1361-6382/acf021} {\path{doi:10.1088/1361-6382/acf021}}.

\bibitem{Battisti:2007jd}
M.~V. Battisti, G.~Montani, {The Big bang singularity in the framework of a generalized uncertainty principle}, Phys. Lett. B 656 (2007) 96--101.
\newblock \href {http://arxiv.org/abs/gr-qc/0703025} {\path{arXiv:gr-qc/0703025}}, \href {https://doi.org/10.1016/j.physletb.2007.09.012} {\path{doi:10.1016/j.physletb.2007.09.012}}.

\bibitem{Battisti:2007zg}
M.~V. Battisti, G.~Montani, {Quantum dynamics of the Taub universe in a generalized uncertainty principle framework}, Phys. Rev. D 77 (2008) 023518.
\newblock \href {http://arxiv.org/abs/0707.2726} {\path{arXiv:0707.2726}}, \href {https://doi.org/10.1103/PhysRevD.77.023518} {\path{doi:10.1103/PhysRevD.77.023518}}.

\bibitem{Battisti:2008rv}
M.~V. Battisti, G.~Montani, {Quantum cosmology with a minimal length}, Int. J. Mod. Phys. A 23 (2008) 1257--1265.
\newblock \href {http://arxiv.org/abs/0802.0688} {\path{arXiv:0802.0688}}, \href {https://doi.org/10.1142/S0217751X08040184} {\path{doi:10.1142/S0217751X08040184}}.

\bibitem{Bosso:2023fnb}
P.~Bosso, O.~Obreg\'on, S.~Rastgoo, W.~Yupanqui, {Black hole interior quantization: a minimal uncertainty approach}, Class. Quant. Grav. 41~(13) (2024) 135011.
\newblock \href {http://arxiv.org/abs/2310.04600} {\path{arXiv:2310.04600}}, \href {https://doi.org/10.1088/1361-6382/ad4fd7} {\path{doi:10.1088/1361-6382/ad4fd7}}.

\bibitem{Ong:2018zqn}
Y.~C. Ong, {Generalized Uncertainty Principle, Black Holes, and White Dwarfs: A Tale of Two Infinities}, JCAP 09 (2018) 015.
\newblock \href {http://arxiv.org/abs/1804.05176} {\path{arXiv:1804.05176}}, \href {https://doi.org/10.1088/1475-7516/2018/09/015} {\path{doi:10.1088/1475-7516/2018/09/015}}.

\bibitem{Battisti:2008qi}
M.~V. Battisti, G.~Montani, {The Mixmaster Universe in a generalized uncertainty principle framework}, Phys. Lett. B 681 (2009) 179--184.
\newblock \href {http://arxiv.org/abs/0808.0831} {\path{arXiv:0808.0831}}, \href {https://doi.org/10.1016/j.physletb.2009.10.003} {\path{doi:10.1016/j.physletb.2009.10.003}}.

\bibitem{Barca:2021epy}
G.~Barca, E.~Giovannetti, G.~Montani, {Comparison of the semiclassical and quantum dynamics of the Bianchi I cosmology in the polymer and GUP extended paradigms}, Int. J. Geom. Meth. Mod. Phys. 19~(07) (2022) 2250097.
\newblock \href {http://arxiv.org/abs/2112.08905} {\path{arXiv:2112.08905}}, \href {https://doi.org/10.1142/S0219887822500979} {\path{doi:10.1142/S0219887822500979}}.

\bibitem{Libermann_1987}
P.~Libermann, Symplectic geometry and analytical mechanics, Dordrecht; Boston: D. Reidel; Norwell, MA: Sold and distributed in the U.S.A. and Canada by Kluwer Academic Publishers, 1987.

\bibitem{Lee_2003}
J.~M. Lee, Introduction to Smooth Manifolds, Springer Science \& Business Media, 2003.

\bibitem{Maggiore:1993kv}
M.~Maggiore, {The Algebraic structure of the generalized uncertainty principle}, Phys. Lett. B 319 (1993) 83--86.
\newblock \href {http://arxiv.org/abs/hep-th/9309034} {\path{arXiv:hep-th/9309034}}, \href {https://doi.org/10.1016/0370-2693(93)90785-G} {\path{doi:10.1016/0370-2693(93)90785-G}}.

\bibitem{Fadel:2021hnx}
M.~Fadel, M.~Maggiore, {Revisiting the algebraic structure of the generalized uncertainty principle}, Phys. Rev. D 105~(10) (2022) 106017.
\newblock \href {http://arxiv.org/abs/2112.09034} {\path{arXiv:2112.09034}}, \href {https://doi.org/10.1103/PhysRevD.105.106017} {\path{doi:10.1103/PhysRevD.105.106017}}.

\bibitem{Gamboa:2000yq}
J.~Gamboa, M.~Loewe, J.~C. Rojas, {Noncommutative quantum mechanics}, Phys. Rev. D 64 (2001) 067901.
\newblock \href {http://arxiv.org/abs/hep-th/0010220} {\path{arXiv:hep-th/0010220}}, \href {https://doi.org/10.1103/PhysRevD.64.067901} {\path{doi:10.1103/PhysRevD.64.067901}}.

\bibitem{Weinstein_1998}
A.~Weinstein, Poisson geometry, Differential Geometry and its Applications 9~(1) (1998) 213–238.
\newblock \href {https://doi.org/10.1016/S0926-2245(98)00022-9} {\path{doi:10.1016/S0926-2245(98)00022-9}}.

\bibitem{PhysRev.71.38}
H.~S. Snyder, \href{https://link.aps.org/doi/10.1103/PhysRev.71.38}{Quantized space-time}, Phys. Rev. 71 (1947) 38--41.
\newblock \href {https://doi.org/10.1103/PhysRev.71.38} {\path{doi:10.1103/PhysRev.71.38}}.
\newline\urlprefix\url{https://link.aps.org/doi/10.1103/PhysRev.71.38}

\end{thebibliography}

\end{document}